\documentclass[12pt]{article}

\usepackage{amssymb}
\usepackage{url}

\newcommand{\rlnabla}{\nabla^{{}^{{}^{\!\!\!\!\!\!\!\!\! \longleftrightarrow}}}}
\newcommand{\be}{\begin{equation}}
\newcommand{\ee}{\end{equation}}
\begin{document}
\begin{titlepage}
\title{Massive spin zero fields in cosmology and the tail-free property}
\author{Valerio Faraoni \\ \\
{\small Department of Physics \& Astronomy, Bishop's University} \\
{\small 2600 College Street, Sherbrooke Qu\'ebec, Canada J1M~1Z7}
}
\date{}
\maketitle   \thispagestyle{empty}  \vspace*{1truecm}
\begin{abstract} 

Fields of spin $s \geq 1/2$ satisfying wave equations in a 
curved space obey the Huygens principle under certain conditions clarified 
by a known theorem. Here this theorem is generalized to spin zero and 
applied to an inflaton field in de Sitter-like space, showing that tails 
of scalar radiation are an unavoidable physical feature.  Requiring the 
absence of tails, on the contrary, necessarily implies an unnatural tuning 
between cosmological constant, scalar field mass, and coupling constant to 
the curvature.

\end{abstract}
\begin{center} 
\end{center}     
\end{titlepage}   \clearpage

\section{Introduction}

There is a long history of studies of massive fields of arbitrary spin 
which satisfy wave equations in curved 
space. These studies have approached the subject from  both the 
mathematical and 
the physical (classical and quantum) sides. An interesting 
aspect of the physics and the mathematics of the propagation of waves 
is the validity, or 
lack thereof, of the Huygens principle. The Huygens principle maintains 
that a delta-like impulse of radiation travels on a sharp front 
propagating at the speed 
of light and reaches an observer all at once. Breaking the Huygens  
principle corresponds, on the contrary,  to~this radiation 
spreading  over a finite spacetime region and arriving to the same 
observer a bit at a time, over an extended period of time. In 
other words, when the Huygens principle is 
violated, the wave propagation is not sharp and it exhibits the phenomenon 
of ``tails'', {\em i.e.}, components of the radiation arriving 
late in time 
in comparison with what one expects from a delta-like source in flat 
3-space and 
propagation along the characteristic surfaces of the corresponding 
hyperbolic partial differential equation.  
There are various possible 
causes of tails, and one of them is the scattering of radiation by  
the spacetime curvature \cite{Hadamard, BakerCopson, 
deWittBrehme, Friedlander}.  Naturally, 
the high frequency modes ({\em i.e.}, those with wavelength much smaller 
than the curvature radius of spacetime)  do not 
``feel'' the spacetime curvature and are essentially unaffected 
by this backscattering phenomenon, which is instead important for 
wavelengths comparable to, or 
larger than, the radius of curvature of spacetime. (We use the 
terminology 
``mode'' and ``wavelength'' but, of course, in general in a curved space 
these are not the usual exponential Fourier modes of flat space.) 
  Unless 
very special conditions are satisfied, 
a field satisfying some wave-like equation and propagating in a curved 
spacetime $\left( {\cal M}, g_{ab} \right)$ will have tails. (Here ${\cal 
M}$ denotes a 
spacetime 
manifold and $g_{ab}$ is a Lorentzian metric on it.) In general, it is 
not known which spacetime geometries and which 
wave equations admit tail-free propagation. Studies of the problem have 
encountered substantial mathematical difficulties, and only partial 
results are available for special geometries and for various wave 
fields (see, {e.g.}, Refs.~\cite{Hadamard, BakerCopson, 
deWittBrehme, Friedlander, Gunther}).

Revisiting older literature, one finds a gap in our knowledge in the wave 
propagation of scalar fields, which are particularly important for 
inflationary cosmology in the early universe and for the late 
accelerated era of a quintessence-dominated universe. 
Specifically, a criterion that must be satisfied by fields of spin $s 
\geq 1/2$ in order not to have tails  was derived  long ago and 
 is generalized here to spin zero fields. We allow these 
scalar fields to couple non-minimally to the Ricci scalar $R$  
as is required, for example in Higgs inflation (see the recent 
review~\cite{Rubio}) and as was considered in 
many earlier inflationary scenarios (\cite{mybook} and references 
therein). We stress the fact that, in general, 
tails are a generic feature of wave 
fields propagating in curved spacetime which is introduced by 
backscattering of the waves off the  curvature. With the physics of wave 
propagation in mind, forbidding tails altogether seems too restrictive 
and, unless 
dictated by some special physical motivation,  conditions 
imposing sharp propagation on a generic curved spacetime are unphysical 
and introduce physical pathologies.

We adopt the metric signature ~$-$~+~+~+ and we use units in 
which 
the speed of light and Newton's constant assume the value unity.  The 
Ricci tensor is computed in terms of the Christoffel symbols 
$\Gamma_{\alpha\beta}^{\delta}$ and its derivatives as 
\be
R_{\mu\rho}= \Gamma^{\nu}_{\mu\rho 
,\nu}-\Gamma^{\nu}_{\nu\rho ,\mu}+ 
\Gamma^{\alpha}_{\mu\rho}\Gamma^{\nu}_{\alpha\nu}- 
\Gamma^{\alpha}_{\nu\rho}\Gamma^{\nu}_{\alpha\mu} 
\ee
and we follow the notation of Ref.~\cite{Waldbook}.

\section{Spin \boldmath{$s\geq 1/2$}}

Let us begin by considering massive fields of spin $s \geq 1/2$ on a 
curved spacetime. This situation was analyzed long ago in 
Ref.~\cite{Wunsch} (see also \cite{Ilge, Ilge2}). This reference proves 
the\\\\
{\em Theorem.} 
A solution of the homogeneous wave equation 
for a massive field of spin 
$s\geq 1/2$ on the spacetime $\left( {\cal M},g_{ab} \right)$ obeys the 
Huygens principle if and only if $\left( {\cal M},g_{ab} \right)$ has 
constant curvature and the Ricci scalar is given by
\be   \label{1}
R=\frac{6m^2}{s} \, ,
\ee
where $m$ is the field mass.\\

The appearance of constant curvature spaces in this theorem is 
particularly relevant for cosmology and for string theories. In fact, 
constant curvature spaces include de Sitter spacetime, which is an 
attractor for early universe inflation in which the cosmos is close (in 
a phase space sense) to a 
de Sitter space, and anti-de Sitter space which is important in string 
theories and for the AdS/CFT correspondence (however, in this article we 
restrict to four spacetime dimensions and, therefore, we lose this 
context which would require extra spatial dimensions).

The exact formulation of the Huygens principle adopted by the author of  
Ref.~\cite{Wunsch} is crucial for the validity of the theorem. The 
Huygens principle holding for solutions of wave equations was already 
formulated  in a clear and physically meaningful way by 
Hadamard \cite{Hadamard} as the 
absence of ``tails'' of radiation generated by an impulsive source. 
However,  several inequivalent definitions have mushroomed over the years,  
including the characteristic propagation property, the progressing-wave 
property, and the tail-free property. These inequivalent definitions are 
 all loosely referred to as ``Huygens' 
principle'', creating considerable confusion in the literature. 
Some order was brought to this area of research by the 
detailed discussion of the relations between  the several  
characterizations of wave tails given in Ref.~\cite{BombelliSonego}.

\section{Spin Zero Fields}

In order to fix the ideas and to generalize the result of 
Ref.~\cite{Wunsch} to the zero spin case relevant for cosmology, 
consider the Klein-Gordon equation for a zero spin field $\phi$,
\be    \label{2}
\square \phi -m^2 \phi -\xi R \, \phi =0\,,
\ee
where 
 $\square \phi \equiv g^{ab}\nabla_a \nabla_b \phi $ is the curved 
spacetime d'Alembertian and the 
dimensionless  constant $ \xi $ describes the direct coupling 
between the field $\phi$ and the Ricci curvature $R$ of 
spacetime. (The reader should keep in mind that we are only 
considering the $s=0$ case here, while Ref.~\cite{Wunsch} studies the more 
complicated wave equations for fields of spin $s\geq 1/2$. Direct 
couplings  
between the field and the curvature tensor or its contractions (such 
as, {e.g.}, those of Refs.~\cite{Waldbook, Tsagas, StarkoCraig}) do  
not appear there.).  

The solutions of Equation~(\ref{2}) can be expressed by a formal Green 
function representation in a normal domain $ \cal{N} $ of spacetime 
that does not contain sources:  
\be \label{3} 
\phi (x) =\int\limits_{\partial 
\cal{N} }dS^{a'}(x') \, G\left( x',x \right) {\rlnabla}_{a'} \, \phi 
(x') \,, 
\ee 
where $\partial \cal{N} $ is the boundary of  $\cal{N}$, $ dS^{a'}(x') $ 
is the oriented volume element on the 
hypersurface  $\partial \cal{N} $ at $ x' $, whereas the operator 
$\rlnabla $  is defined as 
\be \label{4} 
f_{1} \rlnabla f_{2} \equiv f_{1} \nabla f_{2}-f_{2}\nabla f_{1} 
\ee 
for any 
pair of differentiable functions $ \left( f_{1} ,  f_{2} \right)$. As is 
natural, we take 
into account only the retarded Green function $G_R \left( x',x 
\right) $, which is a solution of the wave Equation~(\ref{2}) with an 
impulsive source at the point $x$, so that  
\be \label{5} 
\left[ g^{a'b'}( x') 
\nabla_{a'}\nabla_{b'} -m^2 -\xi R(x')\right] G \left( x',x \right) 
=-\delta \left( x',x \right)\,, 
\ee 
where $ \delta \left( x',x \right) $ is the spacetime Dirac delta  
\cite{BirrellDavies, Weinberg}. 
This is defined such that, for each test function $ f $,
\be \label{6} 
\int d^{4}x' \, \sqrt{-g(x')} \, f(x') \, \delta \left( x',x \right) =f(x)\,. 
\ee 

The retarded Green function $G_R \left( x',x \right) $ can be 
decomposed as  \cite{Hadamard,  BakerCopson, 
deWittBrehme, Friedlander} 
\be \label{7} 
G_R \left( x',x \right)= \Sigma \left( x',x \right)\, \delta_R ( 
\Gamma(x',x) ) + V\left( x',x \right) \, \Theta_R ( -\Gamma(x',x) ) \,,
\ee 
where $\Gamma \left(x',x \right) $ is the square of the proper distance 
between $ x' $ and $ x $ evaluated along the unique geodesic 
that connects  $ 
x' $ and $ x $ in the normal domain $ \cal{N} $ \cite{Friedlander}. In 
our notation $ \delta_R $ and $ 
\Theta_R $ are the Dirac delta  and the 
Heaviside step function which has  support in the past of the 
spacetime point $ x' $. In~a fixed spacetime geometry, the 
coefficient functions $\Sigma$ and $V$ are determined uniquely 
\cite{deWittBrehme,Friedlander}. \mbox{The 
non-vanishing} of $V( x',x)$ corresponds to the presence of wave tails 
propagating {\em inside} the light cone, while the part of the Green 
function proportional to $\Sigma( x',x)$ describes sharp propagation 
along the light cone  \cite{Hadamard, BakerCopson, 
deWittBrehme, Friedlander}. The fact that 
Green functions and propagators of massless fields do not have support 
strictly along the light cone is familiar from quantum field theory in 
curved spacetimes \cite{BirrellDavies, Weinberg}.

In Ref.~\cite{SonegoFaraoni}, an interesting property of the scalar 
field with regard to the Huygens principle 
was reported. By expanding the retarded 
Green function $ G_R( x',x)$ around a point $x$ to approximate the 
four-dimensional curved manifold with its tangent space, it was shown that 
the absence of wave tails, {\em i.e.}, the condition $V \equiv 0$, is 
equivalent to 
\be \label{8} 
m^2+\left( \xi-\frac{1}{6} \right) R(x)=0 \,. 
\ee 

The characterization of the Huygens principle adopted in 
Ref.~\cite{Wunsch} 
actually embodies the requirement that tails be absent. Let us 
discuss what this requirement would amount to for a scalar field  $\phi$ 
relevant in cosmology. By 
imposing this condition on a spin zero field, Equation~(\ref{5}) provides us 
with the condition to be satisfied at the spacetime point $x$ in order 
to have no tails at $x$. If this requirement is extended to {\em every}  
point of spacetime one obtains, as a straightforward consequence of 
Equation~(\ref{8}), the\\\\
{\em Theorem.} The solution of Equation~(\ref{2}) 
in the spacetime $\left( {\cal M}, g_{ab} \right)$ propagates without 
tails 
({\em i.e.}, $V(x',x)=0$ in Equation~(\ref{7}) for any pair of spacetime 
points $x'$ and $x$) if and only if $\left ( {\cal M}, g_{ab} \right)$ is 
a constant curvature space~and
\be   \label{9}
R=\frac{6m^2}{1-6\xi} \, .
\ee

A special case is given by the parameter value $\xi=1/6$, for which  
Equation~(\ref{8}) implies that 
there are no tails of the scalar field if and only if $m=0$, regardless  
of the curvature.  As is well known \cite{Waldbook}, if~$\xi=1/6$ and 
$m=0$, then Equation~(\ref{2}) is conformally 
invariant \cite{Waldbook} and the absence of tails for a free scalar 
field propagating in empty spacetime carries over to  (anti-)de Sitter 
space, which has constant curvature and is conformally flat, like all 
Friedmann-Lema\^itre-Robertson-Walker spaces \cite{Waldbook}.

For $s=0$, the case $\xi=0$ in which the scalar field couples minimally to 
the curvature mimics the equations studied in Ref.~\cite{Wunsch}, and 
there is some 
analogy with the theorem proven in Ref.~\cite{Wunsch}. Most~likely, the 
inclusion of non-minimal coupling terms between the wave field and 
 the Riemann tensor or its contractions in the wave equations of 
\cite{Wunsch} would modify the condition~(\ref{1}).

In general relativity and for $\xi 
\neq 1/6$, the spacetime metric $g_{ab}$ which determines sharp 
propagation of the scalar field $\phi$ at all spacetime points is 
obtained by imposing that a cosmological constant $\Lambda$ be present 
(which ensures that the Ricci 
curvature $R$ does not depend on the spacetime point) and that its 
value is related with the 
scalar field mass and with the coupling constant by 
\be \label{10} 
\Lambda=\frac{R}{4}=\frac{3m^2}{2(1-6\xi )} \,. 
\ee 

Therefore, by imposing tail-free 
propagation (in the sense described above), one necessarily obtains the de 
Sitter space if 
$\xi >1/6$ or the anti-de Sitter 
space if $\xi < 1/6$. These two constant curvature spaces are extremely 
important  for cosmology and for string theories and the AdS/CFT 
correspondence, respectively. Requiring the absence of 
tails, which would be interesting from the mathematical point of view,  
would necessarily imply the unnatural tuning 
relation~(\ref{10}) between cosmological constant $\Lambda$, scalar field 
mass $m$, and 
coupling constant $\xi$, which cannot be justified on a physical basis.

It is interesting that a relation similar to Equation~(\ref{10}) appears in 
theories in which gravity is described by a non-linear $\sigma$-model, and 
in 
massive gravity in particular. Ref.~\cite{Kodama} studied the propagation 
of the massive graviton in this type of theory in anti-de Sitter space, 
and derived a relation similar to Equation~(\ref{10}) between the graviton mass 
$m$, 
the cosmological constant $\Lambda$, and another parameter $\alpha$  
of the theory. The graviton mass $m$ is certainly a necessary parameter 
in the study of the validity or violation of the Huygens principle in 
this 
class of theories. To be more precise, the theory is described by the 
action \cite{Kodama, MPLA, EPL}
\be
S=\frac{1}{16\pi } \int d^4 x \sqrt{-g} \left[ R +m^2 U\left( g, \phi 
\right) \right] \,, \label{Saction1}
\ee
where the effective potential $U$ contains two free parameters 
$\alpha_{3,4}$ \cite{Kodama, MPLA, EPL}, 
\be
 U\left( g, \phi \right) = U_2 +\alpha_3 U_3 +\alpha_4 U_4 
\,.\label{Saction2}
\ee 
and a relation similar to Equation~(\ref{10}) involving $m, \Lambda$, and 
$\alpha 
\equiv 1+3\alpha_3$ emerges in anti-de Sitter space (see Equation~(19) of 
Ref.~\cite{Kodama}). The two free parameters play a role similar to 
that of the coupling parameter $\xi$ in our discussion. In the 
theory~(\ref{Saction1}) the vacuum 
can be single or degenerate, depending on the relation between these  two 
free parameters. In this sense, these free parameters (resp., 
$\xi$) control the 
background vacuum 
experienced by the massive graviton (resp. scalar) as it propagates in the 
constant curvature anti-de Sitter (resp., de Sitter) background.  This 
interesting coincidence will be explored in the future.

\section{Conclusions}

Thus far, we have restricted our considerations  
to the mathematical aspects of the propagation of a scalar field in a 
curved space. At this point, it is appropriate to remember the physical 
reasons for the occurrence of tails. 
The possible causes are \cite{BombelliSonego, SonegoFaraoni}:

\begin{itemize}

\item The presence of a mass term ($m\neq 0$) or of  a potential 
$V(\phi)$ in the wave equation (here we restrict the potential to a mass 
term $V=m^2\phi^2/2$ that 
keeps the Klein-Gordon equation linear). As~an 
example, the wave-like solutions of the Klein-Gordon Equation~(\ref{1}) in 
four-dimensional flat spacetime $\left( \mathbb{R}^4, \eta_{ab} 
\right)$ exhibit tails whenever $m\neq 0$.

\item The spacetime dimension is another important factor for the presence 
or absence of tails. For~example, in $d$-dimensional Minkowski spacetime,  
the solutions of 
Equation~(\ref{2})  have tails for odd 
$d$, but not for even $d$ \cite{BakerCopson,BarrowTipler}. Here, however,  
we limit ourselves to considering four spacetime dimensions.

\item Scattering of the waves off the background spacetime 
curvature---this is the situation in which the most interesting physics 
appears.

\end{itemize}

The mathematical literature has focussed on looking for the conditions and 
geometries for which tails of radiation are absent, presumably because 
sharp propagation without tails seems a desirable feature in physics and 
in other areas of research. Indeed, sharp propagation is required for 
transmitting efficiently information carried by wave signals. For example, 
the absence of tails in the transmission of electric signals along nerves 
is necessary for the functioning of complex neural networks and for the 
existence of sophisticated organisms, and it has been used as an anthropic 
argument to discriminate the dimensionality of space (In general, 
the Huygens principle is violated by any linear partial differential 
equation in which the solution depends on an odd number of variables 
\cite{Hadamard, BakerCopson, Garabedian}. For even number of variables (as 
in a four-dimensional spacetime), the Huygens principle may or may not 
hold, which may be difficult to assess. The tail-free property can be seen 
as being equivalent to the Huygens principle~\cite{BakerCopson, 
BombelliSonego,BarrowTipler}. It is natural, therefore, that 
there is renewed interest about tails of radiation in curved space, and in 
cosmology in particular, by the information theory community 
\cite{Martinez1, Martinez2, Martinez3, Martinez4, Stefano, Simidzija, 
StarkoCraig}).

Studying the various conditions for the absence of wave tails of  
massive fields of arbitrary spin (as in Ref.~\cite{Wunsch}) is certainly 
legitimate and well motivated from the mathematical point of view. 
However, it~is not 
easily justifiable from the physical point of view. In fact, a field with 
mass $m\neq 0$  has tails due the fact that it is massive (or that it 
has a potential $V(\phi)$, which amounts to a field-dependent mass) and to 
the backscattering off the background curvature of spacetime. The absence 
of tails for such a field means that the two effects cancel out, and this  
cancellation is completely unphysical. From the point of 
view of field theory, this 
situation would correspond to the case of a field $\phi$ with non-zero 
mass which 
propagates locally with the speed of light. This rather fictional, and 
clearly fine-tuned, phenomenon goes against our experience in flat space 
and does not have any experimental support. The case, contemplated above, 
of a massive scalar field with coupling constant  $\xi \neq 1/6$ 
travelling in de Sitter or anti-de 
Sitter space is an example in which this pathological behaviour occurs at 
{\em every} point of spacetime. In the light of these considerations, one 
concludes that a wave tail is indeed a natural, and even desiderable, 
feature for a massive field of any spin in four spacetime dimensions. 
Therefore, imposing the absence of tails (as done in Ref.~\cite{Wunsch}) 
may have considerable mathematical interest but is not significant, nor 
realistic, from the physical point of view. Indeed, scalar wave tails may 
have effects that are important for cosmological applications 
\cite{cosmoconse, cosmoconse2, cosmoconse3, cosmoconse4} and are relevant 
for the propagation of super-horizon modes and nearly horizon-sized modes 
of an inflaton field. Moreover, the violation of the Huygens principle has 
consequences for relativistic quantum communication, as pointed out in 
recent studies \cite{Martinez1, Martinez2, Martinez3, Martinez4}.  Tails 
of gravitational (spin~two) radiation emitted from compact objects have 
also been studied \cite{gwtails1, gwtails1bis, gwtails2, gwtails3, 
gwtails4} in relation with the {\em LIGO} and {\em VIRGO} laser 
interferometric detectors which have recently discovered gravitational 
waves \cite{LIGO, LIGO2, LIGO3}, although in that case tails are a higher 
order effect in the wave amplitude and, therefore, not of immediate 
interest \cite{gwtails1, gwtails1bis, gwtails2, gwtails3, gwtails4}. For 
physical applications, it seems that only spacetimes and wave fields which 
have non-pathological behaviour with respect to wave tails are admissible 
(by non-pathological we mean that a true field mass can never be removed  
by a tail which is caused by scattering off the curvature of spacetime) 
\cite{SonegoFaraoni, SonegoFaraoniJMP}.

In Ref.~\cite{SonegoFaraoni} it was argued that, to ensure that tails of a 
scalar field are present when $m\neq 0$ and to preclude physical 
pathologies, only the value $\xi=1/6$ of the coupling constant between 
$\phi$ and $R$ should be allowed.  This result was later confirmed in 
Ref.~\cite{GribPoberii}, and its consequences for inflationary cosmology 
and for the late, quintessence-dominated universe, were explored in 
Refs.~\cite{FaraoniPRD, FaraoniPLA, Faraonipeyresq, mybook, present1, 
present2, present3, present4}. In the future, it will be interesting to 
explore the generalization of the wave equations obtained by including 
explicit couplings between fields of various spins and the Riemann tensor 
$R_{abcd}$ or its contractions ({e.g.}, \cite{Waldbook, Tsagas, 
StarkoCraig}).  Perhaps the imposition that there are no causal 
pathologies such as massive fields propagating exactly on the light cone 
will fix the values of the coupling constants (perhaps, again, to nonzero 
values), as it happens in the $s=0$ case \cite{SonegoFaraoni}. In the case 
of spin~$1/2$, this approach may potentially be of interest for neutrinos 
emitted in supernova collapse or for the cosmological neutrino background 
in the early universe.

\section*{Acknowledgments} 

We thank two referees for useful comments. This work is supported, in 
part, by the Natural Sciences and Engineering Research Council of Canada 
(Grant No. 2016-03803).

\end{document}